\documentclass[pra,amsmath,amssymb, twocolumn, showpacs]{revtex4}

\usepackage{dcolumn}% Align table columns on decimal point

\usepackage{bm}% bold math

\usepackage[dvips]{graphicx}

\usepackage{epsfig}

\begin{document}

\title{Geometric phases and Bloch sphere constructions for SU(N), with a complete description of SU(4)}

\author{D.\ Uskov and A.\  R.\  P. Rau$^{*}$}
\affiliation{Department of Physics and Astronomy, Louisiana State University,
Baton Rouge, Louisiana 70803-4001, USA}

%\date{\today}

\begin{abstract}

A two-sphere (``Bloch" or ``Poincare") is familiar for describing the dynamics of a spin-1/2 particle or light polarization. Analogous objects are derived for unitary groups larger than SU(2) through an iterative procedure that constructs  evolution operators for higher-dimensional SU in terms of lower-dimensional ones. We focus, in particular, on the SU(4) of two qubits which describes all possible logic gates in quantum computation. For a general Hamiltonian of SU(4) with 15 parameters, and for Hamiltonians of its various sub-groups so that fewer parameters suffice, we derive Bloch-like rotation of unit vectors analogous to the one familiar for a single spin in a magnetic field. The unitary evolution of a quantal spin pair is thereby expressed as rotations of real vectors. Correspondingly, the manifolds involved are Bloch two-spheres along with higher dimensional manifolds such as a four-sphere for the SO(5) sub-group and an eight-dimensional Grassmannian manifold for the general SU(4). This latter may also be viewed as two, mutually orthogonal, real six-dimensional unit vectors moving on a five-sphere with an additional phase constraint.  

\end{abstract}


\maketitle

\section{Introduction: The Bloch sphere and its extension}
In the study of the dynamics of a spin-1/2 particle, a visual metaphor that has played a powerful role is that of the ``Bloch sphere" \cite{Schleich}. Pure states of the system are represented by the tip of a vector from the origin to the surface of such a unit sphere S$^2$. In the field of nuclear magnetic resonance (nmr) \cite{nmr} and elsewhere, transformations between states are then viewed as rotations of that vector, described by the Bloch equation of motion, $\dot {\vec m} = -2 {\vec B} \times {\vec m}$, for a magnetic moment in a magnetic field ${\vec B}$. Thus, various sequences of nmr manipulations can be pictured in a nice geometrical way as successive rotations, and this has now become central to our intuition of spin dynamics. The relevant group of unitary transformations is SU(2), a rank-one, three-parameter group that is the double covering group of the three-dimensional rotation group SO(3) \cite{groups}. The three operators of angular momentum, $(J_x, J_y, J_z)$, are the generators of these groups. A canonical set of parameters of SO(3) are the Euler angles. Integer values $j=0, 1, \ldots $ provide various $(2j+1)$-dimensional  representations, while for SU(2), the half-odd integers occur as well. The two coordinates on S$^2$, together with a phase, provide the three parameters describing the full state. 

This latter phase is often not accessible as, for instance, when dealing with the density matrix $\rho$. Mixed states also are naturally accommodated in this picture. They are represented by points inside the sphere so that the vector is of length less than unity. Correspondingly, Tr $\rho ^2 < {\rm Tr}\, \rho$, which constitutes a definition of a mixed state \cite{mixedstate}. States of light polarization, also a two-valued object, map onto the same mathematics and geometry through the ``Poincare" sphere \cite{Poincare}.

It would be of interest to have analogous geometrical pictures for multiple spins, especially in today's fields of quantum computation, cryptography, and teleportation, because the fundamental elements of these subjects are built up of a few qubits \cite{Nielsen}. Thus, all logic gates for quantum computation can be built up from qubit pairs, while teleporting one qubit state requires an entangled pair held by the sender and receiver, for a total of three qubits. With SU(2$^p$) being the relevant group for $p$ qubits, this calls for a similar geometrical description of higher SU($N$). In this paper, we develop such a picture, through an easily accessible procedure which iteratively descends from $N$ to $N-n$, with $n < N$, in a manner that closely follows the description of SU(2). 

Our procedure also applies when $N$ is odd, a situation that does not arise with qubits but elsewhere widely in physics (for example, qutrits \cite{qutrit}, neutrino oscillations \cite{neutrino}, the quark model and quantum chromodynamics (QCD), etc.) The $N \times N$ matrices of Hamiltonians and evolution operators are viewed as built up of $2 \times 2$ block matrices through this $N=(N-n)+n$ decomposition, the block matrices then described in terms of the Pauli spinors of SU(2). Each step of this iterative reduction introduces an analog of the Bloch sphere, albeit of higher dimension and more complex structure, and constructs the effective Hamiltonians of dimension $(N-n)$ and $n$ for the next step. In this manner, using no more than the operations familiar from the SU(2) case, the full construction for SU($N$) is achieved. 

The philosophy behind such a construction may be seen as generalizing Schwinger's philosophy for representations of SU(2) or SO(3), where higher $j$-representations are constructed from those of the fundamental, $j=1/2$ \cite{Schwinger}. We now do the analogous step of using SU(2) as the template for solving larger SU($N$). In particular, for the important case of SU(4) for two qubits, we give a complete description of the manifolds and phases involved and analytical expressions for them. Note again, as with light polarization and spin-1/2, that the mathematics of $N$-level systems in quantum optics, atomic and molecular physics, and elsewhere, is the same as that we describe in the language of multiple qubits. This provides an even wider context for our results.

The arrangement of this paper is as follows. Section II describes the basic iterative decomposition of the evolution operator for SU($N$), mimicking the familiar procedure for spin-1/2.  With $N=4$, and $n=2$, Section III specializes the results to SU(4), the case of two qubits, when all the manipulations involved are in terms of Pauli spinors. It also applies these results to Hamiltonians involving a restricted set of operators of the full group. An interesting one is SO(5), which can be described by a $5 \times 5$ antisymmetric matrix that is the analog of the $3 \times 3$ antisymmetric one for the magnetic field ${\vec B}$ in the Bloch equation. Section IV then considers Hamiltonians requiring the full SU(4) group for their description. Linear equations, analogous to the Bloch equation, are derived in terms of vectors $\vec {m}$, five- and six-dimensional vectors, respectively, for the SO(5) and full SU(4) cases. The latter also correspond to so-called ``Pl\"{u}cker coordinates" \cite{Plucker} which are also presented. Appendix A deals with the generalization to non-Hermitian Hamiltonians, and Appendix B presents the isomorphism between SU(4) and the groups Spin(6) and SO(6) which we exploit.

\section{Iterative construction of evolution operator in $N$ dimensions}

We wish to obtain the evolution operator ${\bf U}^{(N)}(t)$ for the $N$-dimensional time-dependent Hamiltonian ${\bf H}^{(N)}$:

\begin{equation}
{\bf H}^{(N)}(t) =\left(
\begin{array}{cc}
{\bf H}^{(N-n)}(t) & {\bf V}(t) \\
{\bf V}^{\dagger}(t) & {\bf H}^{(n)}(t)
\end{array}
\right).
\label{eqn1}
\end{equation}
We have blocked the Hamiltonian into $(N-n)$- and $n$-dimensional blocks, the diagonal blocks being square matrices while the off-diagonal ${\bf V}$ is $(N-n) \times n$ and ${\bf V}^{\dagger}$ is $n \times (N-n)$. Although our discussion is for Hermitian ${\bf H}^{(N)}$, the procedure can also apply more generally, in which case the off-diagonal blocks will not be simply related as adjoints (see Appendix A). We will also assume ${\bf H}^{(N)}$ to be traceless, again a restriction that can be easily relaxed, the time integral of the trace becoming an overall phase of ${\bf U}^{(N)}$.

To solve the evolution equation, with an over-dot denoting derivative with respect to time,

\begin{equation}
i\dot {\bf U}^{(N)}(t) =   {\bf H}^{(N)}(t){\bf U}^{(N)}(t), \,\, {\bf U}^{(N)}(0) ={\bf I},
\label{eqn2}
\end{equation}
we similarly block the unitary matrix, writing it also as a product of three factors, the first two further grouped as $\tilde{\bf U}_1$ and the second, $\tilde{\bf U}_2$, block-diagonal in form:

\begin{eqnarray}
{\bf U}^{(N)}(t) & = & \tilde{\bf U}_1 \tilde{\bf U}_2, \,\,\,\, \tilde{\bf U}_1= e^{{\bf z}(t) A_{+}} e^{{\bf w}^{\dagger} (t)A_{-}}, \nonumber \\
\tilde{\bf U}_1 & = & \left(
\begin{array}{cc}
{\bf I}^{(N-n)} & {\bf z}(t) \\
{\bf 0}^{\dagger} & {\bf I}^{(n)}
\end{array} \right) \left(
\begin{array}{cc}
{\bf I}^{(N-n)} & {\bf 0} \\
{\bf w}^{\dagger}(t) & {\bf I}^{(n)}
\end{array} \right), \nonumber \\
\tilde{\bf U}_2 & = &  \left(
\begin{array}{cc}
\tilde{\bf U}^{(N-n)} (t) & {\bf 0} \\
{\bf 0}^{\dagger} & \tilde{\bf U}^{(n)} (t)
\end{array} \right),
\label{eqn3}
\end{eqnarray}
where $A_{\pm}$ are matrix generalizations of the Pauli spin step-up/down $\sigma _{\pm}$, and ${\bf z}$ and ${\bf w}^{\dagger}$ are rectangular matrices of complex parameters.

The above structure, with $\tilde{\bf U}_1$ having blocks of zero in the lower and upper off-diagonal blocks of its matrix factors, is crucial in our method. For the case of spin-1/2 and SU(2), the form of a product of three factors, each an exponentiation of one of the Pauli spinors, is well known \cite{groups}. Their Cartesian form, with Euler angles in the exponents, is the familiar choice but we choose instead the triplet, $(\sigma _{\pm}, \sigma_z)$, when the first two factors have zero off-diagonal entries. This introduces complex ${\bf z}$ and ${\bf w}^{\dagger}$ in place of the Euler angles, and makes the individual factors in Eq.~(\ref{eqn3}) not separately unitary although our construction ensures unitarity of the full ${\bf U}^{(N)}(t)$. Further, for non-Hermitian $H$ when $U$ is non-unitary, our construction still applies. The specific structure of an upper and lower triangular matrix and a diagonal one proves fruitful, giving simpler equations for ${\bf z}$ and ${\bf w}^{\dagger}$, which will have at most quadratic nonlinearity in these parameters and not more complicated  trigonometric dependences as with the Euler angle decomposition \cite{Rau98, Uskov}. They also yield more naturally to a geometrical picture of the manifolds they describe.

A remark about notation. We will use the symbol tilde when the corresponding Hamiltonians or evolution operators may not be Hermitian or unitary, respectively. Unitarity of the full ${\bf U}^{(N)}(t)$ leads to relations between ${\bf z}$ and ${\bf w}^{\dagger}$ which would otherwise be independent for evolution under a non-Hermitian Hamiltonian (see Appendix A),

\begin{eqnarray}
{\bf z} = - {\bf w}\, {\mbox {\boldmath $\gamma$}_2} & = & {-\mbox {\boldmath $\gamma$}_1} \,{\bf w}, \nonumber \\
{\mbox {\boldmath $\gamma$}_1} \equiv  \tilde{\bf U}^{(N-n)} \tilde{\bf U}^{(N-n)\dagger} & = & {\bf I}^{(N-n)} +{\bf z}{\bf z}^{\dagger}, \nonumber \\
{\mbox {\boldmath $ \gamma$}_2}^{-1} \equiv \tilde{\bf U}^{(n)} \tilde{\bf U}^{(n)\dagger} & = & ({\bf I}^{(n)} +{\bf z}^{\dagger} {\bf z})^{-1}.
\label{eqn4}
\end{eqnarray}

With ${\bf U}=\tilde{\bf U}_1 \tilde{\bf U}_2$, Eq.~(\ref{eqn2}) formally reduces to the evolution of $\tilde{\bf U}_2$ alone with an effective Hamiltonian \cite{Uskov, Heff},

\begin{equation}
i\dot{\tilde{\bf U}}_2 = \tilde {\bf H}_{\rm eff} \tilde{\bf U}_2, \,\,\tilde{\bf H}_{\rm eff}=\tilde{\bf U}_1^{-1}{\bf H}\tilde{\bf U}_1 -i\tilde{\bf U}_1^{-1}\dot{\tilde{\bf U}}_1.
\label{eqn5}
\end{equation}
A key element of our construction lies in this effective Hamiltonian and corresponding evolution for the reduced problem. Since $\tilde{\bf U}_2$ and this equation are block diagonal, the off-diagonal blocks in ${\bf H}_{\rm eff}$ on the right-hand side must vanish. This condition leads to the defining equation for ${\bf z}$,

\begin{equation}
i\dot{\bf z}={\bf H}^{(N-n)} {\bf z} + {\bf V}-{\bf z}({\bf V}^{\dagger}{\bf z}+{\bf H}^{(n)}).
\label{eqn6}
\end{equation}

For SU(2), when $N=2, n=1$, all the matrices above reduce to single numbers and Eq.~(\ref{eqn6}) is a Riccati equation for the complex $z$. More generally, it is a matrix Riccati equation \cite{Reid}, and its solutions are involved in the subsequent construction. With the off-diagonal blocks of Eq.~(\ref{eqn5}) accounted for, the diagonal ones defining the Hamiltonians for the $(N-n)$ and $n$ problems remain, and are given by $({\bf H}^{(N-n)} -{\bf z}{\bf V}^{\dagger})$ and $({\bf H}^{(n)}+{\bf V}^{\dagger}{\bf z})$, respectively. Although the overall trace is preserved in our construction and remains zero, these individual Hamiltonians are neither traceless nor Hermitian. The equations for ${\bf z}$ need to be solved numerically in general but form a smaller set than the $N^2$ elements in the original Eq.~(\ref{eqn2}).

To set up the process for iteration, the above individual Hamiltonians in $(N-n)$- and $n$-dimensional subspaces must be rendered Hermitian and traceless. The latter is easily achieved, by subtracting Tr $({\bf H}^{(N-n)} -{\bf z}{\bf V}^{\dagger})$ and Tr $({\bf H}^{(n)}+{\bf V}^{\dagger}{\bf z})$ from them. These traces being equal and opposite, this translates into the introduction of a phase, the integral of the trace, in ${\bf U}^{(N)}$, representing a relative phase between the two subspaces. 

There are alternative methods for rendering the Hamiltonians Hermitian, the most accessible one being through 

\begin{equation}
\tilde{\bf U}_1^{\dagger} \tilde{\bf U}_1 =  \left(
\begin{array}{cc}
{\mbox {\boldmath $ \gamma$}_1}^{-1} & {\bf 0} \\
{\bf 0}^{\dagger} & {\mbox {\boldmath $ \gamma$}_2}
\end{array} \right)
\equiv \left(
\begin{array}{cc}
{\bf g}_1 {\bf g}_1^{\dagger} & {\bf 0} \\
{\bf 0}^{\dagger} & {\bf g}_2 {\bf g}_2^{\dagger}
\end{array} \right)^{-1}.
\label{eqn7}
\end{equation}
The first part of this equation is the observation that $\tilde{\bf U}_1^{\dagger} \tilde{\bf U}_1$ is block diagonal. This suggests the second part of the equation, namely, the definition of an inverse through two ``Hermitian square-root" matrices ${\bf g}_i$. Together, they serve as a gauge factor to unitarize according to

\begin{equation}
{\bf U}_1 =\tilde{\bf U}_1 \left(
\begin{array}{cc}
{\bf g}_1 & {\bf 0} \\
{\bf 0}^{\dagger} & {\bf g}_2
\end{array} \right).
\label{eqn8}
\end{equation} 
With that, the second factor, $\tilde{\bf U}_2$, in Eq.~(\ref{eqn3}) is also unitarized,

\begin{equation}
{\bf U}_2= \left(
\begin{array}{cc}
{\bf g}_1^{-1} & {\bf 0} \\
{\bf 0}^{\dagger} & {\bf g}_2^{-1}
\end{array} \right)\tilde{\bf U}_2.
\label{eqn9}
\end{equation}

After some algebra, the explicitly Hermitian forms of the two diagonal block Hamiltonians of dimension $(N-n)$ and $n$ are 

\begin{eqnarray}
{\bf H}^{(N-n)}\!\!\! & =\! \!\!& \frac{i}{2} [\frac{d}{dt} {\bf g}_1 ^{-1}, {\bf g}_1]\! +\!\! \frac{1}{2}\! \left( {\bf g}_1^{-1} ({\bf H}^{(N-n)}\!\!-{\bf z}{\bf V}^{ \dagger})\, {\bf g}_1 \!\! +\!{\rm hc} \right)\!, \nonumber \\
\!\!{\bf H}^{(n)}\!\!\! & =\!\!\! & \frac{i}{2} [\frac{d}{dt} {\bf g}_2 ^{-1}, {\bf g}_2 ]  \!\!+\!\! \frac{1}{2}\! \left( {\bf g}_2 ^{-1} ({\bf H}^{(n)}\!\!+{\bf z}^{\dagger}{\bf V}) \, {\bf g}_2 \!\! + \! {\rm hc} \right)\!,
\label{eqn10}
\end{eqnarray}
with commutator brackets in the first term, and hc in the second term denoting the Hermitian conjugate of the preceding expression. Again, the trace of each Hamiltonian in Eq.~(\ref{eqn10}) can be subtracted to render them traceless; as clear by inspection, this is the same trace discussed just above. These Hamiltonians in Eq.~(\ref{eqn10}) can now be treated further as SU($N-n$) and SU($n$) problems.

The ${\mbox {\boldmath $\gamma$}}$ matrices in Eq.~(\ref{eqn4}) are Hermitian with non-negative eigenvalues because of their origin from $\tilde{\bf U}_1^{\dagger} \tilde{\bf U}_1$. This permits their decomposition into ${\bf g}$ as shown in Eq.~(\ref{eqn7}). The ${\bf g}$ matrices and their inverses in Eq.~(\ref{eqn7})-Eq.~(\ref{eqn10}), are square roots of them, and because any power, including fractional ones, are Hermitian term by term in a formal power-series expansion, we can choose ${\bf g}$ also as Hermitian. The use of identities such as

\begin{equation}
{\bf z}^{\dagger} {\mbox {\boldmath $\gamma$}_1}^p={\mbox {\boldmath $\gamma$}_2}^p {\bf z}^{\dagger}, \,\,\, {\mbox {\boldmath $\gamma$}_1}^p {\bf z}={\bf z} {\mbox {\boldmath $\gamma$}_2}^p, \,\,
\label{eqn11}
\end{equation}
serves to express all ${\bf g}$ in terms of the linearly independent set of matrices of dimension $(N-n)$ or $n$, whichever is smaller. With $n=2$, this means that all the algebra of calculating such square-root matrices and the subsequent evaluation of the effective Hamiltonian in Eq.~(\ref{eqn10}) reduces to manipulation of Pauli matrices.       

A count of the parameters is instructive. The original SU($N$) evolution involves $(N^2-1)$ elements and, therefore, grows quadratically with $N$. These are divided in the above construction into the $2n(N-n)$ elements in ${\bf z}$, which for small $n$ grows only linearly with $N$. The rest are contained in the elements of the SU($N-n$) and SU($n$) and the single phase between those two subspaces. Our construction of higher SU($N$) evolution in terms of smaller ones, with the template in Eq.~(\ref{eqn3}) of three factors as in SU(2), resembles the Schwinger scheme of generating higher $j$ representations of SU(2) or SO(3) from the fundamental one of $j=1/2$ \cite{Schwinger}. Whereas that scheme was for higher representations but of the same group, SU(2), our procedure extends in the direction of larger groups SU($N$).

In mathematical language of base manifolds and fiber bundles \cite{fiber}, the SU(2) and its Bloch sphere are seen as the bundle [SU(2)/U(1)] $\times $ U(1), the former the two-sphere S$^2$ base and the latter U(1) phase the fiber. Likewise, our construction is in terms of the base manifold [SU($N$)/(SU($N-n)$ $\times $ SU($n$) $\times $ U(1))] and the fiber (SU($N-n)$ $\times $ SU($n$) $\times $ U(1)). For SU(2), there is a single complex $z$ that defines the base manifold. The Bloch sphere of a unit three-dimensional vector $\vec{m}$ corresponding to $z$ is then constructed by inverse stereographic projection from R$^2$ to S$^2$. Similar structures of a $\vec{m}$ associated with the larger ${\bf z}$ will be considered in the next sections. 

\section{The case of SU(4), with application to its sub-groups}
An important case is of $N=4$. Four-level systems are commonly considered in quantum optics and molecular systems and, of course, in today's quantum computation where they describe two qubits \cite{Nielsen}. Since all logic gates can be built up from such qubit pairs, the study of the evolution operator for such $N=4$ problems is of current interest. As a combined description of spin and isospin, SU(4) also has central importance in the study of nuclei and particles \cite{Close}. The group also occurs in the description of unusual magnetic phases of $f$ electron states in CeB$_6$ \cite{Ohkawa}. Both choices $n=1,2$ in the general procedure of Section II lead to interesting decompositions, with the latter the more natural for qubit applications. We now turn to this case. 

In physics terms, a 4-level Hamiltonian has three real parameters along the diagonal to fix the energy positions of the levels. (One overall element, represented by the trace, can be subsumed as an uninteresting definition of the zero energy reference level, leading also to an irrelevant overall phase in the evolution operator.) In addition, six off-diagonal couplings, which are complex, make for a total of 15 parameters to describe the full Hamiltonian. Symmetries often reduce this number so that the Hamiltonian involves only a smaller number as a closed sub-algebra. For two identical qubits, there are indeed such symmetries which reduce the number of independent energies and couplings of a 4-level system.

With $N=4, n=2$, all the matrices involved in the previous section can be rendered in terms of Pauli spinors and the unit $2 \times 2$ matrix. A general Hamiltonian of SU(4) has 15 independent operators and time-dependent parameters multiplying them. A standard, explicit rendering of the 15 $4 \times 4$ matrices is given in \cite{Rau2, Hilbers}. ${\bf z}$ comprises four complex quantities, $(z_4, z_i)$, and the matrix Riccati equation reduces to coupled first-order equations in them with quadratic nonlinearity. Deferring this general case to the next section, we consider first the smaller sets of operators of various sub-groups of SU(4). 

{\it su(2) $\times$ su(2) sub-algebra:} Consider first a Hamiltonian consisting of only six of the 15 operators. Since our construction is representation independent, in a suitable representation, the six may be viewed as two independent, mutually commuting, triplets that obey su(2) algebra. Clearly, each then may be expected to have its own geometrical description in terms of a Bloch sphere and phase. In our above, general formulation, this result is realized as follows. Thus, consider two independent magnetic moments, characterized by the standard Pauli matrices $\sigma$, in time-varying magnetic fields $A(t)$ and $B(t)$ which may also be independent, with Hamiltonian $H = \vec{\sigma}^{(1)} \cdot {\vec A} + \vec{\sigma}^{(2)} \cdot {\vec B}$. Using a standard set of $ 4 \times 4$ matrices \cite{Rau2} to cast this Hamiltonian in the form of Eq.~(\ref{eqn1}), we have ${\bf V}=(A_x-iA_y){\bf I}$ and ${\bf H}^{(1,2)}= \vec{\sigma} \cdot {\vec B} \pm A_z {\bf I}$. The ${\bf z}$ in Eq.~(\ref{eqn3}) also reduces, as with ${\bf V}$, to a unit operator with a single complex coefficient $z_4$ obeying a Riccati equation in Eq.~(\ref{eqn6}). The gamma matrices in Eq.~(\ref{eqn4}) are also proportional to the unit operator, thus simplifying Eq.~(\ref{eqn10}), the ${\bf g}$ dropping out. As a result, the Hermitian matrices in the block-diagonal effective Hamiltonian take the form of the same $\vec{\sigma} \cdot {\vec B}$ plus/minus a term proportional to a unit matrix. The first term is viewed as for a single spin with a Bloch sphere and a phase, the second represents a phase between the two $2 \times 2$ spaces. The complex $z_4$ can again be inverse stereographically projected into another two-sphere as in the Bloch construction. We arrive, therefore, at the same initial expectation, that a simultaneous viewing in terms of two Bloch vectors in individual two-spheres, along with their fibers, provides the geometrical picture for all such qubit-pair systems. A specific physical example occurs in the construction of optimal quantum NOT operations \cite{Novotny}.

{\it su(2) $\times$ su(2) $\times$ u(1) sub-algebra:} Another sub-algebra, involving seven of the 15 operators, has been considered before \cite{Rau2, Ganesh}. It has the symmetry of SU(2) $\times $ SU(2) $\times $ U(1). In a suitable representation, such a Hamiltonian can be cast as a diagonal form in Eq.~(\ref{eqn1}) plus a term which is proportional to the unit operator in both diagonal blocks but with equal and opposite sign. Such an operator commutes with all the other six, themselves comprised of two mutually commuting triplets of $4 \times 4$ matrices \cite{Rau2}. With ${\bf V}=0$, ${\bf z}$ also vanishes and we reduce trivially to the two independent SU(2) and a phase between the two spaces, together accounting for the 7 parameters of this problem. An example is provided by the CNOT gate constructed with two Josephson junctions \cite{Nakamura}. Many such sets of seven operators, one of which commutes with all the remaining six, have been identified through a general procedure in footnote 11 of \cite{Ganesh}.

{\it so(5) sub-algebra:} Proceeding further to other sub-groups, a non-trivial example is provided by a $H$ that involves ten operators satisfying an so(5) sub-algebra of su(4). Again, there are many such sets of ten operators/matrices which close under commutation within the full set of 15 as noted in footnote 11 of \cite{Ganesh}. As a physical example, a four-level system of two symmetric pairs, as naturally so with two identical qubits, has only two real parameters along the diagonal in its $H$. Selection rules often restrict the off-diagonal coupling between the levels from six to four, thus introducing four complex, or eight real, parameters. The net result of such symmetric four-level systems is a ten-parameter problem \cite{Legare}. Such $H$ fall into this so(5) sub-algebra. The corresponding group is the so-called spin group Spin(5) which is the double-covering group of SO(5), the group of five-dimensional rotations, much as Spin(3), isomorphic to SU(2), is the covering group of SO(3) \cite{spin}. All such Spin(5) or SO(5) will themselves have a Spin(4) or SO(4) sub-group, which in turn has the two mutually commuting SU(2) or SO(3) discussed above so that the ten matrices can be conveniently viewed as two sets of commuting triplets plus four more which transform like a four-dimensional vector under SO(4). For completeness here in this paper, we briefly summarize results on this so(5) sub-algebra that were published elsewhere \cite{Uskov}; see also \cite{Rangan}. 

In a convenient representation that uses Pauli matrices for two spins \cite{Rau2}, we have $H(t)=F_{21}\sigma^{(2)}_z-F_{31}\sigma^{(2)}_y+F_{32}\sigma^{(2)}_x-F_{4i}\sigma^{(1)}_z\sigma^{(2)}_i+F_{5i}\sigma^{(1)}_x\sigma^{(2)}_i-F_{54}\sigma^{(1)}_y$, where the ten arbitrarily time-dependent coefficients $F_{\mu \nu}(t)$ form a $5 \times 5$ antisymmetric real matrix. (We will use $\mu,\nu=1-5$ and $i,j,k=1-3$ and summation over repeated indices.) Several quantum optics and multiphoton problems of four levels driven by time-dependent electric fields have such a Hamiltonian. It has also been considered extensively in coherent population transfer in many molecular and solid state systems \cite{Legare}.  Casting this Hamiltonian in the form of Eq.~(\ref{eqn1}), we have

\begin{equation}
{\bf H}^{(1,2)}=(\mp F_{4k} -\frac{1}{2} \epsilon_{ijk} F_{ij}) \sigma_k, {\bf V}=iF_{54} {\bf I}^{(2)}+F_{5i} \sigma_i.
\label{eqn12}
\end{equation}

With the matrix Riccati equation in Eq.~(\ref{eqn6}) cast in terms of Pauli spinors together with coefficients $ z_{\mu}=z_4,z_i$: ${\bf z}=z_4 {\bf I}^{(2)} -iz_i \sigma_i$, it takes the form 

\begin{equation}
\dot{z}_{\mu}=F_{5\mu}(1-z_{\nu}^2)+2F_{\mu \nu}z_{\nu}+2F_{5\nu}z_{\nu}z_{\mu}.
\label{eqn13}
\end{equation}
(As an alternative, ${\bf V}$ and $ {\bf z}$ can also be rendered in terms of quaternions $(1,-i\sigma_i)$.) ${\mbox {\boldmath $ \gamma $}_1}$ and ${\mbox {\boldmath $ \gamma $}_2}$ in Eq.~(\ref{eqn4}) become equal and proportional to a unit matrix, $(1+z_{\mu}z_{\mu}){\bf I}^{(2)}$. The structure of Eq.~(\ref{eqn13}) admits to the four quantities $z$ being real. The effective Hamiltonian in Eq.~(\ref{eqn10}) in terms of these $z$ becomes

\begin{equation}
{\bf H}_{\rm eff}^{(1,2)}={\bf H}^{(1,2)}-\epsilon_{ijk} z_i F_{5j} \sigma_k \mp F_{5j} z_4\sigma_j \pm F_{54 }z_i \sigma_i.
\label{eqn14}
\end{equation}

We can now construct a five-dimensional unit vector $\vec{m}$ out of the four $z$,

\begin{equation}
m_{\mu}=\frac{-2z_{\mu}}{(1+z_{\nu}^2)}, \,m_5=\frac{(1-z_{\nu}^2)}{(1+z_{\nu}^2)},\,\,\,\, \mu,\nu=1-4. 
\label{eqn15}
\end{equation}
The nonlinear Eq.~(\ref{eqn6}), or Eq.~(\ref{eqn13}) in $z$, becomes of simple, linear Bloch-like form, 

\begin{equation}
\dot{m}_{\mu}=2F_{\mu\nu}m_{\nu},\,\,\,\,\, \mu,\nu=1-5.
\label{eqn16}
\end{equation}

As in the single spin case, this represents an inverse stereographic projection, now from the four-dimensional plane $z \in R^4$ to the four-sphere $S^4$. It provides a higher-dimensional polarization vector for describing such two spin problems. With ${\bf z}$ so described, the two effective SU(2) Hamiltonians in Eq.~(\ref{eqn14}), when solved in turn, give the complete solution. In all, such Hamiltonians possessing Spin(5) symmetry are, therefore, described by the geometrical picture of one S$^4$ and two S$^2$ spheres along with two phases.

{\it su(3) sub-algebra:} Four-level systems with only two independent energy parameters along the Hamiltonian's diagonal and three complex off-diagonal couplings constitute a su(3) sub-algebra with 8 parameters. A general three-level system, embedded into four with the fourth level completely uncoupled, constitutes a trivial example of such an su(3) sub-algebra but less trivial examples can also occur. The ${\bf z}$ now has two non-zero complex $z$ for a total of four parameters. The description of this four-dimensional manifold, as well as the remaining SU(2) and a U(1) phase, parallel the discussion of the general SU(4) in the next section, and will be presented elsewhere \cite{Sai}. Therefore, we omit details except to note that setting $z_4=-iz_1$ and $z_3=-iz_2$ in Section IV reduces to such a SU(3) symmetry. 

\section{The general SU(4) Hamiltonian involving all fifteen operators}

Instead of the Hamiltonians considered in Section III which involve sub-algebras of the full two-qubit system, consider an arbitrary $4 \times 4$ Hamiltonian with its entire complement of 15 operators/matrices. Such a $H$ is obtained by adding to the previous Spin(5) Hamiltonian considered above the five additional terms, $F_{65} \sigma ^{(1)}_z +F_{64} \sigma ^{(1)}_x +F_{6i} \sigma ^{(1)}_y \sigma ^{(2)}_i$. Correspondingly, Eq.~(\ref{eqn11}) gets an additional term $\pm F_{65} {\bf I}^{(2)}$ in the diagonal ${\bf H}^{(1,2)}$ while in ${\bf V}$, the $F_{5 \mu}$ are replaced by $F_{5 \mu} -iF_{6 \mu}$. Thus, the full SU(4) amounts to a simple modification of the previously considered Spin(5) by adding a term proportional to the unit operator to the diagonal blocks and making the four $F_{5 \mu}$ complex, with $F_{6 \mu}$ absorbed as their imaginary parts.             

The Riccati Eq.~(\ref{eqn13}), now for complex $z$, becomes

\begin{eqnarray}
\dot{z}_{\mu} & = & F_{5\mu}(1-z_{\nu}^2) -iF_{6 \mu}(1+z_{\nu}^2)+2F_{\mu \nu}z_{\nu} \nonumber \\
 \!\!& + &\!\! 2(F_{5\nu}+iF_{6\nu})z_{\nu}z_{\mu}-2iF_{65}z_{\mu},  \mu,\nu=1-4.
\label{eqn17}
\end{eqnarray}
The two gammas in Eq.~(\ref{eqn4}) are given by

\begin{eqnarray}
{\mbox {\boldmath $\gamma$}_{1,2}} & = &  (1+z_{\mu}^2){\bf I}^{(2)} +i(z_i^{*}z_4-z_4^{*}z_i)\sigma_i \nonumber \\
 & \pm & \frac{1}{2} i \epsilon_{ijk} (z_i z_j^{*}-z_j z_i^{*}) \sigma_k.
\label{eqn18}
\end{eqnarray}
Their square-root matrices $g_{1,2}$ can also be evaluated in terms of the Pauli matrices and the two SU(2) effective Hamiltonians then constructed in explicitly traceless and Hermitian form.

Just as the very structure of Eq.~(\ref{eqn13}) suggests that $z_{\mu}$ and $(1-z_{\nu}^2)$ with suitable normalization define a five-dimensional unit vector $\vec{m}$ in Eq.~(\ref{eqn15}), the occurrence of $z_{\mu}, (1-z_{\nu}^2), (1+z_{\nu}^2)$ in Eq.~(\ref{eqn17}) suggests now the introduction of six quantities according to

\begin{equation}
m_{\mu}=\frac{-2z_{\mu}}{De^{i\phi}}, \,m_5=\frac{(1-z_{\nu}^2)}{De^{i\phi}},\, m_6=-i\frac{(1+z_{\nu}^2)}{De^{i\phi}},
\label{eqn19}
\end{equation}
with 

\begin{eqnarray}
D & \equiv & (1+2|z_{\nu}|^2+z_{\mu}^2 {z^{*}_{\nu}}^2)^{1/2}, \nonumber \\
\dot{\phi} & = & \!\!-2F_{65}+iF_{5\mu}(z^{*}_{\mu}\!-\!z_{\mu})+F_{6\mu}(z^{*}_{\mu}\!+\!z_{\mu}).
\label{eqn20}
\end{eqnarray}

As with the so(5) case in Section III, with such a set of six complex quantities $\vec{m}$, the nonlinear Riccati equation for the four complex $z_{\mu}$ in Eq.~(\ref{eqn17}) becomes a linear Bloch-like equation as before,

\begin{equation}
\dot{m}_{\mu}=2F_{\mu\nu}m_{\nu}, \,\,\,\mu,\nu=1-6.
\label{eqn21}
\end{equation}
Once again, the $m_{\mu}$ obey a first-order equation with an antisymmetric matrix which describes rotations. Since the 15 $F_{\mu\nu}$ are real, the real and imaginary parts of the six $m_{\mu}$ each obey such a rotational transformation. These six-dimensional rotations reflect the isomorphism between the groups SU(4) and SO(6) (more accurately, its covering group  Spin(6)) and suggest a mapping between their generators (see Appendix B). 

To get a geometrical picture of the manifold $m$, we note first the relations, 

\begin{equation}
m_{\mu}^2=0, \,\, |m_{\mu}|^2=2,
\label{eqn22}
\end{equation}
which amount to three constraints. In addition, only the derivative, not the value, of $\phi$ is determined in Eq.~(\ref{eqn20}). Thereby, the number of independent parameters in $m_{\mu}$ is eight just as in the complex $z_{\mu}$, themselves built from ${\bf z}$. The description of such an eight-dimensional manifold will be taken up in the next sub-section but we note here the reduction to the previous so(5) example. This follows upon setting $F_{65}=0, F_{6\mu}=0$ which makes $\phi =0$ and $D=(1+z_{\nu}^2)$ in Eq.~(\ref{eqn20}), and reduces $m_{\mu}$ and $m_5$ to the values in Eq.~(\ref{eqn15}) whereas $m_6=-i$. This, of course, makes $\vec{m}$ a five-dimensional unit vector and its manifold the four-sphere S$^4$. The first relation in Eq.~(\ref{eqn22}), of the vanishing of a square, hints at Grassmannian elements, to be discussed further below. 

\subsection{Nature of the manifold describing $(z,m)$ for general SU(4)}

Our construction of the evolution operator for $(N=4, n=2)$ in Eq.~(\ref{eqn3}) is in terms of the eight-dimensional base manifold ${\bf z}$ and a fiber consisting of two residual SU(2) along its diagonal blocks and a U(1) phase between them: SU(4) $\rightarrow$ [SU(4)/SU(2) $\times$ SU(2) $\times$ U(1)] $\times$ [SU(2) $\times$ SU(2) $\times$ U(1)]. To describe the former base manifold, consider first [SU(4)/SU(2) $\times$ SU(2)], which is a nine-dimensional manifold. It can also be described in terms of spin-groups as Spin(6)/Spin(4). The six complex $m_{\mu}$ in Eq.~(\ref{eqn18}) with the three constraints in Eq.~(\ref{eqn22}) constitute such a manifold called a Stiefel manifold {\bf St}(6, 2, {\sf R}) $\cong \Re ^9$, this name being given to manifolds consisting of $n$ orthogonal vectors from an $N$-dimensional space $\Re^N$ \cite{Stiefel}. Geometrically, the second relation in Eq.~(\ref{eqn22}) states that the real and imaginary parts of $m$ are six-dimensional unit vectors while the first relation expresses their mutual orthogonality. Therefore, one can view the manifold as a five-sphere S$^5$ with another four-sphere S$^4$ attached at each point on it. The absolute value of the phase parameter $\phi$ in Eq.~(\ref{eqn19}) and Eq.~(\ref{eqn20}) being undefined, reduces such a manifold by one dimension to [SU(4)/SU(2) $\times$ SU(2) $\times$ U(1)], which is equivalent to the reduction from the Stiefel to a Grassmannian manifold {\bf G}(4, 2, {\sf C}) according to {\bf St}(6, 2, {\sf R}) $\cong {\bf G}(4, 2, {\sf C}) \times $ U(1). Such a Grassmannian manifold, which has eight dimensions, thereby describes the ${\bf z}$ in Eq.~(\ref{eqn3}) or its equivalent $z_{\mu}$ in Eq.~(\ref{eqn17}) or $m_{\mu}$ in Eq.~(\ref{eqn19}).       

A more accessible geometrical picture is to consider a single five-sphere S$^5$ embedded in six-dimensional space and two six-dimensional unit vectors from the origin to the surface to represent the real and imaginary parts of $m_{\mu}$. The two vectors are always taken as orthogonal, so that one views such an orthogonally-coupled pair rotating within the sphere \cite{circle}. This nine-dimensional object, combined with the zero reference of $\phi$ being undefined, is our eight-dimensional manifold of interest.

\subsection{Description in Pl\"{u}cker coordinates}

An alternative view of these manifolds is provided in terms of what are termed Pl\"{u}cker coordinates, defined as a set of six complex parameters $(P_{12}, P_{13}, P_{14}, P_{23}, P_{24}, P_{34})$ formed as minors of the $2 \times 4$ sub-matrix of the last two columns of an arbitrary, unitary SU(4) matrix \cite{Plucker},

\begin{equation}
{\bf U}=\left( 
\begin{array}{cccc}
u_{11} & u_{12} & u_{13} & u_{14} \\ 
u_{21} & u_{22} & u_{23} & u_{24} \\ 
u_{31} & u_{32} & u_{33} & u_{34} \\
u_{41} & u_{42} & u_{43} & u_{44}
\end{array}
\right).
\label{eqn23}
\end{equation}
They obey the relations

\begin{equation}
P_{12}P_{34}-P_{13}P_{24}+P_{14}P_{23} =0, \,\, \sum |P_{ij}|^2=1.
\label{eqn24}
\end{equation}
They are combinations of the $m_{\mu}$ according to 

\begin{equation}
\left( 
\begin{array}{c}
P_{12} \\ 
P_{13} \\ 
P_{14} \\
P_{23} \\
P_{24} \\
P_{34}
\end{array}
\right) = \frac{1}{2} \left(
\begin{array}{c}
im_6 -m_5 \\
im_1 +m_2 \\
-im_3 +m_4 \\
-im_3 -m_4 \\
-im_1 +m_2 \\
im_6 +m_5
\end{array} \right).
\label{eqn25}
\end{equation}

The linear equations for $m_{\mu}$ in Eq.~(\ref{eqn20}) translate into a similar linear equation 

\begin{equation}
i\dot{\bf P}={\bf H}{\bf P}, \,\, {\bf P} \equiv (P_{12},-P_{13}, P_{14}, P_{23}, P_{24}, P_{34}),
\label{eqn26}
\end{equation}
with

\begin{equation}
{\bf H}_P \! =\!\! \left( 
\begin{array}{cccccc}
H_{11,22} & H_{41} & H_{31} & -H_{42} & H_{32} & 0 \\ 
H_{14} & H_{11,33} & -H_{34} & -H_{12} & 0 & -H_{32} \\ 
H_{13} & -H_{43} & H_{11,44} & 0 & H_{12} & H_{42} \\
-H_{24} & -H_{21} & 0 & H_{22,33} & H_{34} & -H_{31} \\
H_{23} & 0 & H_{21} & H_{43} & H_{22,44} & -H_{41} \\
0 & -H_{23} & H_{24} & -H_{13} & -H_{14} & H_{33,44}
\end{array}
\right),
\label{eqn27}
\end{equation}
where we have adopted the notation for the diagonal entries: $H_{ii,jj}=H_{ii}+H_{jj}$.

Actually, the above equations for ${\bf P}$ can be arrived at directly from the evolution equation $i\dot{\bf U}={\bf H}{\bf U}$ because the elements of ${\bf P}$ are quadratic in the elements of ${\bf U}$ in Eq.~(\ref{eqn23}): $P_{ij}=i \varepsilon_{ijkl}u^{(3)}_k u^{(4)}_l$, and $i\dot{u}^{(3)}_k=H_{kj}u^{(3)}_j$. Also, ${\bf z}$ can be defined in terms of the two minors on the right in Eq.~(\ref{eqn23}):

\begin{equation}
{\bf z} = \left(
\begin{array}{cc}
u_{13} & u_{14} \\
u_{23} & u_{24}
\end{array} \right) /
\left( 
\begin{array}{cc}
u_{33} & u_{34} \\
u_{43} & u_{44}
\end{array} \right),
\label{eqn28}
\end{equation} 
the matrix in the denominator assumed to be non-singular. Writing $U$ in Eq.~(\ref{eqn23}) in the form in Eq.~(\ref{eqn3}), the first factor $\tilde{U}_1$ involving ${\bf z}$ is a map of the Grassmannian manifold {\bf G}(4, 2, {\sf C}) onto ${\sf C}^4$, and provides a partial coordinization of that manifold. Elements of {\bf G}(4, 2, {\sf C}) are two-dimensional complex hyperplanes spanned by vectors ${\bf u}_3 =(u_{13}, u_{23}, u_{33}, u_{43})^T$ and ${\bf u}_4 =(u_{14}, u_{24}, u_{34}, u_{44})^T$. The Pl\"{u}cker coordinates provide a unique identification of such planes. They are an analog of the coordinization of the $n$-dimensional sphere S$^n$ by an $(n+1)$-dimensional unit vector $\vec{m}$ as in Section III.

The matrix ${\bf H}_P$ in Eq.~(\ref{eqn27}) being Hermitian, ${\bf P}^{\dagger}{\bf P}$ = constant = 1. This can be verified by the relation between $P$'s and $m$'s in Eq.~(\ref{eqn25}) which involves a unitary matrix so that ${\bf P}^{\dagger}{\bf P}=\frac{1}{2} m^{\dagger}m$, and combining with Eq.~(\ref{eqn22}). Further, a symplectic structure can be introduced. 
Defining a $6 \times 6$ matrix ${\bf \Omega} \equiv {\bf \delta}_{i,7-j}$ with non-zero entries of 1 only along the anti-diagonal, the first relation in Eq.~(\ref{eqn24}) can be rendered as ${\bf P}^T {\bf \Omega} {\bf P}=0$, and the matrix ${\bf H}_P$, a generator of the symplectic group {\bf Sp}(6, \sf C),

\begin{equation}
{\bf H}_P {\bf \Omega} + {\bf \Omega}{\bf H}_P^T = {\rm Tr} ({\bf H}_P){\bf \Omega}=0. 
\label{eqn29}
\end{equation}
Any two vectors ${\bf P}_i$, evolving according to Eq.~(\ref{eqn26}), satisfy ${\bf P}_1^T (t) {\bf \Omega} {\bf P}_2 (t)$ =constant. If ${\bf P}_1 {\bf \Omega} {\bf P}_2 =0$, then the two hyperplanes defined by ${\bf P}_i$ intersect, and if $|{\bf P}_1 {\bf \Omega} {\bf P}_2| =1$, they do not.

Geometrically, the set ${\bf P}^{\dagger}{\bf P}=1$ is a sphere S$^{11}$, the algebraic relation ${\bf P}{\bf \Omega}{\bf P}=0$ determining a 9-dimensional sub-manifold, an intersection between S$^{11}$ and the affine variety of roots of the polynomial equation ${\bf P}{\bf \Omega}{\bf P}=0$. This manifold may be denoted $\Re ^9$. Multiplication by a phase acts as a transformation group on this manifold, that is, if ${\bf P} \in \Re ^9$, then ${\bf P}e^{i\phi} \in \Re ^9$. Therefore, {\bf G}(4, 2, {\sf C}) is a quotient space $\Re ^9/$U(1) and has eight dimensions. The connection to SU(4) is, as noted before, $\Re ^9 \cong $SU(4)/(SU(2) $\times$ SU(2)) $\cong$ Spin(6)/ Spin(4). The stability sub-group of a vector ${\bf P} \in \Re ^9$ is SU(2) $\times$ SU(2) while the stability sub-group of $\Re ^9$/U(1) is SU(2) $\times$ SU(2) $\times$ U(1). Since Spin(6)/Spin(5) $\cong$ S$^5$ and Spin(5)/Spin(4) $\cong$ S$^4$, we can identify the fibration of $\Re ^9$ with S$^5 \times$ S$^4$.

\section{Summary}

We have presented a complete analysis of the evolution operator for SU($N$), setting up its construction in a hierarchical way in terms of those for smaller SU($N-n$) and SU($n$), with $n<N$ and arbitrary. The evolution operator is written as a product of two $N \times N$ matrices, the second of which is block diagonal in $(N-n) \times (N-n)$ and $n \times n$ of the smaller groups. The first factor is obtained through a ${\bf z}$, which is an $(N-n) \times n$ complex matrix obeying a matrix Riccati equation. Its solutions determine both the first factor as well as the Hermitian matrices for the subsequent $N-n$ and $n$ evolution problems. 

This general constructive method is applied especially to a four-level system with special emphasis on two qubits. The general symmetry is of SU(4), a 15-parameter group. Our procedure expresses the evolution operator as a product of two 4 $\times$ 4 matrices, the second of which is block diagonal, each block an SU(2) problem. The ${\bf z}$ is also a $2 \times 2$ matrix with complex entries in general and obeys a matrix Riccati equation. Alternatively, we transform ${\bf z}$ into a six-dimensional complex vector $\vec{m}$, whose real and imaginary parts both separately undergo linear, six-dimensional rotational transformations. This is exactly analogous to the linear Bloch equation for real three-dimensional rotations of a vector to represent the evolution operator for a single spin in a magnetic field.  

Just as a Bloch sphere describes the three-dimensional vector $\vec{m}$ for a single spin (and, together with a phase, the complete SU(2)), we also present the geometrical manifold describing ${\bf z}$ or its equivalent six-dimensional complex vector $\vec{m}$. Together with two residual SU(2) problems and a phase, this provides a complete description of the quantum evolution operator for SU(4). For certain sub-algebras of SU(4), the manifold is an analogous higher-dimensional sphere; a four-sphere, for example, for an so(5) sub-algebra. For the most general SU(4), we have an eight-dimensional Grassmannian manifold. We provide a picture of it as two five-spheres with an orthogonality and phase constraint. These geometrical objects may serve for all possible four-level and two qubit systems the useful purpose that the Bloch sphere has for two-level and single qubit problems in physics. 

\section*{APPENDIX A: EXTENSION TO NON-UNITARY EVOLUTION FOR A NON-HERMITIAN HAMILTONIAN}

The iterative method of Section II for the evolution operator in Eq.~(\ref{eqn2}) through writing it as in Eq.~(\ref{eqn3}) applies also when $H$ in Eq.~(\ref{eqn1}) is not Hermitian and, therefore, the evolution not unitary. However, ${\bf z}$ and ${\bf w}$ in Eq.~(\ref{eqn3}) are no longer simply related as in Eq.~(\ref{eqn4}) but obey independent equations, the former still in Riccati form but the latter given in terms of ${\bf z}$. Thus, instead of Eq.~(\ref{eqn1}), consider

\begin{equation}
\tilde{\bf H}^{(N)}(t) =\left(
\begin{array}{cc}
\tilde{\bf H}^{(N-n)}(t) & {\bf V}(t) \\
{\bf Y}^{\dagger}(t) & \tilde{\bf H}^{(n)}(t)
\end{array}
\right),
\label{eqnA1}
\end{equation}
where we have again indicated by tildes non-Hermiticity, and ${\bf V}$ and ${\bf Y}$ are not equal but independent.

Writing $\tilde{\bf U}^{(N)}(t)$ again as in Eq.~(\ref{eqn3}), Eq.~(\ref{eqn6}) now becomes

\begin{eqnarray}
i\dot{{\bf z}} & = & (\tilde{\bf H}^{(N-1)}{\bf z} -{\bf z}\tilde{\bf H}^{(n)} -{\bf z}{\bf Y}^{\dagger}{\bf z} +{\bf V}, \nonumber \\
i\dot{{\bf w}}^{\dagger}\!\! & = & \!\! {\bf w}^{\dagger}({\bf z}{\bf Y}^{\dagger}\!\!-\tilde{\bf H}^{(N-1)} \!) \!+(\!\tilde{\bf H}^{(n)}\!-{\bf Y}^{\dagger}{\bf z}){\bf w}^{\dagger}\!\!+\! {\bf Y}^{\dagger}.
\label{eqnA2}
\end{eqnarray}
The residual problems of $(N-n)$ and $n$ dimension then become

\begin{equation}
\left( \begin{array}{cc}
i\dot{\tilde{\bf U}}^{(N-n)} \! & \!\! {\bf 0} \\
{\bf 0}^{\dagger} \! & \!\! i\dot{\tilde{\bf U}}^{(n)}
\end{array} \right) \! = \! \left(
\begin{array}{cc}
\tilde{\bf H}^{(N-1)} -{\bf z} {\bf Y}^{\dagger} \! & \!\! {\bf 0} \\
{\bf 0}^{\dagger} \! & \!\! \tilde{\bf H}^{(n)} + {\bf Y}^{\dagger}{\bf z}
\end{array} \right) \tilde{\bf U}_2.
\label{eqnA3}
\end{equation}

\section*{APPENDIX B: DESCRIPTION OF EVOLUTION AS SIX-DIMENSIONAL ROTATIONS}

For a single spin or qubit, the rewriting of the quantum evolution operator, which is complex, as rotational transformations of a real, unit vector in three dimensions given by the Bloch equation, rests on the isomorphism of the group SU(2) to SO(3) (or its double covering Spin(3)). A similar isomorphism between the groups SU(4) and SO(6) (or its extension Spin(6)) underlies the construction in Sections III and IV of the complex evolution operator for two qubits in terms of rotations of a vector in six dimensions. Both groups are described by 15 real parameters through an antisymmetric $F_{\mu\nu}, \mu, \nu=1, 2, \ldots 6$. In Sections III and IV, explicit expressions are given for the Hamiltonian with each of these parameters multiplying one of the 15 complex generators of SU(4) in a standard representation of Pauli matrices, $\vec{\sigma}^{(1)} \otimes \mathcal{I}^{(2)}, \mathcal{I}^{(1)} \otimes \vec{\sigma}^{(2)}, \vec{\sigma}^{(1)} \otimes \vec{\sigma}^{(2)}$. An alternative rendering in terms of the 15 generators of SO(6) is useful and recorded here.

The Hamiltonian in Section IV, apart from a factor of $\frac{1}{2}$, can be cast in terms of a matrix array

\begin{equation}
\!\!\left(\!\! 
\begin{array}{cccccc}
\!\!0 & \!\!\sigma^{(2)}_z & \!\!-\sigma^{(2)}_y & \!\!-\sigma^{(1)}_z \sigma^{(2)}_x & \!\!\sigma^{(1)}_x \sigma^{(2)}_x & \!\!\sigma^{(1)}_y\sigma^{(2)}_x \\ 
\!\!-\sigma^{(2)}_z & \!\!0 & \!\!\sigma^{(2)}_x & \!\!-\sigma^{(1)}_z \sigma^{(2)}_y & \!\!\sigma^{(1)}_x \sigma^{(2)}_y & \!\!\sigma^{(1)}_y \sigma^{(2)}_y \\ 
\!\!\sigma^{(2)}_y & \!\!-\sigma^{(2)}_x & \!\!0 & \!\!-\sigma^{(1)}_z \sigma^{(2)}_z & \!\!\sigma^{(1)}_x \sigma^{(2)}_z & \!\!\sigma^{(1)}_y \sigma^{(2)}_z \\
\!\!\sigma^{(1)}_z \sigma^{(2)}_x & \!\!\sigma^{(1)}_z \sigma^{(2)}_y & \!\!\sigma^{(1)}_z \sigma^{(2)}_z & \!\!0 & \!\!-\sigma^{(1)}_y & \sigma^{(1)}_x \\
\!\!-\sigma^{(1)}_x \sigma^{(2)}_x & \!\!-\sigma^{(1)}_x \sigma^{(2)}_y & \!\!-\sigma^{(1)}_x \sigma^{(2)}_z & \!\!\sigma^{(1)}_y & \!\!0 & \!\!\sigma^{(1)}_z\\
\!\!-\sigma^{(1)}_y \sigma^{(2)}_x & \!\!-\sigma^{(1)}_y \sigma^{(2)}_y & \!\!-\sigma^{(1)}_y \sigma^{(2)}_z & \!\!-\sigma^{(1)}_x & \!\!-\sigma^{(1)}_z & \!\!0
\end{array}\!\!
\right),
\label{eqnB1}
\end{equation}
which is explicitly anti-symmetric. We thus have $H=2F_{\mu\nu}L_{\nu\mu}$. Analogous to the familiar triplet of angular momentum generators, six-dimensional generators of SO(6) are given by $L_{\mu\nu} =-il_{\mu\nu}$, where the $l$ are 15 real antisymmetric $6 \times 6$ matrices with only two non-zero entries, $+1$ in the $(\mu\nu)$ and $-1$ in the $(\nu\mu)$ position:

\begin{equation}
(l_{\mu\nu})_{\rho\sigma} = \delta_{\mu\rho} \delta_{\nu\sigma} - \delta_{\mu\sigma} \delta_{\nu\rho}.
\label{eqnB2}
\end{equation}
Their commutators close:

\begin{equation}
[l_{\mu\nu}, l_{\rho\sigma}]=\delta_{\nu\rho}l_{\mu\sigma} +\delta_{\mu\sigma} l_{\nu\rho} -\delta_{\nu\sigma} l_{\mu\rho} -\delta_{\mu\rho} l_{\nu\sigma},
\label{eqnB3}
\end{equation}
so that $L_{\mu\nu}$ form an so(6) algebra.

The array in Eq.~(\ref{eqnB1}) is also a convenient display of the generators of the various sub-groups of SO(6) of lower-dimensional rotations. Either upper left or lower right corner $2 \times 2$ blocks describe the SO(2) generator of one of the qubits. Adding a third row and column gives the full triplet of SO(3) generators. To this can be added a next row and column of three non-zero entries to give the six generators of SO(4). For this purpose, any of the three remaining row/column can be employed, each giving an SO(4), the three added entries transforming as a vector under SO(3). This continues. Adding another row and column's four new entries, which transform as a vector under SO(4) (further subdividing into three components that transform as a vector and one as a scalar under the previous SO(3)), gives the ten SO(5) generators. The final sixth row/column adds five entries, an SO(5) vector, to give the full 15 generators of SO(6). This hierarchical nesting of SO sub-groups, together with the corresponding Clifford structure with Pauli matrices in Eq.~(\ref{eqnB1}), accounts for the richness of the structures in the isomorphic groups SU(4) and SO(6), one we have exploited in Sections III and IV. Note that the linear Bloch-like equation for $\vec{m}$ in Eq.~(\ref{eqn21}) for a general SU(4) Hamiltonian reduces to the same antisymmetric form for its sub-groups such as in Eq.~(\ref{eqn16}), all the way down to the standard Bloch equation for a single qubit, whose SO(3) antisymmetric $F_{ij}$ is usually written as a vector product with a magnetic field.


\begin{thebibliography}{}
\bibitem{spin} The SO(n) orthogonal group of rotations is familiar and suffices for most of our purposes, although technically it is the Spin(n) groups which are the double covering groups that occur as sub-groups of SU(n) with the Clifford structure of the Pauli matrices. Even more accurately, in the sense of the Cartan decomposition, it is SU(2) $\otimes$ SU(2)/$Z_2$, where $Z_2$ is a group of $\pm 1$, which is isomorphic to SO(4) whereas SU(2) $\otimes$ SU(2) is isomorphic to Spin(4).
\bibitem{circle} A lower dimensional analog is instructive. Two real quantities, $(x_1, x_2)$ with $x_1^2+x_2^2=1$ define a circle or one-sphere S$^1$. Two complex numbers $(z_1, z_2)$, with $z=x+iy$, obeying the constraints in Eq.~(\ref{eqn22}), $|z_i|^2=1, z_i^2=0$, also define a one-dimensional manifold of two real, orthogonal, unit vectors $(\vec{x}, \vec{y})$. The $m_{\mu}$ in Eq.~(\ref{eqn19}) and Eq.~(\ref{eqn22}) are the six-dimensional extension of this.
\end{thebibliography}
\end{document}